\documentclass[letterpaper,twocolumn,english,aps,prl,groupedaddress,superscriptaddress,showpacs,amsfonts]{revtex4}

\usepackage{epsfig}
\usepackage{graphicx}
\usepackage{dcolumn}
\usepackage{bm}

\DeclareMathAlphabet{\mathpzc}{OT1}{pzc}{m}{it}


\newcommand{\beq}{\begin{equation}}
\newcommand{\eeq}{\end{equation}}
\newcommand{\bea}{\begin{eqnarray}}
\newcommand{\eea}{\end{eqnarray}}

\begin{document}

\title{Phase Diagram of the Holstein-Hubbard Two-Leg Ladder}

\author{Ka-Ming Tam}

\affiliation {Department of Physics, Boston University, 590 Commonwealth Ave., Boston, MA 02215}

\author{S.-W. Tsai}

\affiliation{Department of Physics, University of California, Riverside, CA 92507}

\author{D.~K.~Campbell}

\affiliation {Department of Physics, Boston University, 590 Commonwealth Ave., Boston, MA 02215}

\author{A.~H.~Castro Neto}

\affiliation {Department of Physics, Boston University, 590 Commonwealth Ave., Boston, MA 02215}

\date{\today}

\begin{abstract}
Using a functional renormalization group method, we obtain the phase diagram of the two-leg ladder system within the Holstein-Hubbard 
model, which includes both electron-electron and electron-phonon interactions.  Our renormalization group technique allows us to analyze the problem for both weak and strong electron-phonon 
coupling. We show that, in contrast results from conventional 
weak coupling studies, electron-phonon interactions can dominate electron-electron interactions because of retardation effects. 
\end{abstract}
\maketitle

The renormalization group (RG) method has provided
fundamental  understanding of the stable phases of matter in terms of the basin of  attraction of fixed points (FP). Quantum phase transitions (QPT) between  different fixed points can be described in terms of the stability of these FP relative to the RG flow. In the theory of metals, Landau's Fermi liquid theory can be described either as a stable FP
of the renormalization group flow under repulsive interactions, or as an
unstable FP under attractive interactions to a  superconducting state (SC)  described by BCS theory \cite{Shankar}.  Apart from determining the stability of a FP,  the RG method also provides direct information about the energy scales (gaps)  and length scales (correlation lengths) associated with a QPT. Hence, the  RG not only provides qualitative information, but also quantitative  information, about the different phases of matter.

\begin{figure}[phase]
\centerline{
\includegraphics*[height=0.22\textheight,width=0.40\textwidth,viewport=45 215
510 575,clip]{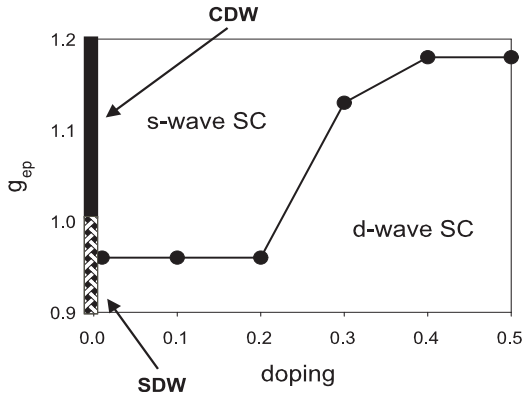}} \caption{Phase diagram of Holstein-Hubbard ladder
as a function of electron-phonon coupling $g_{\rm{ep}}$
and doping ($t=1$,$U=2$, $\omega_0=1$). Zero doping corresponds to
half-filling and doping of 0.5 corresponds to quarter-filling.}
\label{phd}
\end{figure}

In practice, however, there are very few examples of problems in which the RG can be implemented beyond weak coupling \cite{kondo}. 
Furthermore, another
well-known challenge to the RG has been its implementation in
problems with retardation, such as the electron-phonon (e-ph) problem,  because
 of the introduction of multiple, frequency-dependent interaction channels that
 renormalize in a collective, energy dependent, way.  At weak e-ph coupling, 
when retardation plays a small role, the RG can be implemented using 
approximations such as the so-called "two-step" RG \cite{Zimanyi,Seidel}. 
Usually, the two-step RG leads to an enhancement of the instabilities that 
exist when only electron-electron (e-e) interactions are present, but does not
describe phases where retardation effects due to the e-ph coupling become 
dominant over e-e effects. In fact, we show that new phases, such as charge 
density wave (CDW) and s-wave SC, emerge at strong e-ph coupling 
(see Fig.~\ref{phd}).

Recently, the RG was extended to study the non-perturbative physics of the e-ph problem in the large $N$ limit (where $N = E_F/\Lambda$, $E_F$ is the Fermi 
energy and $\Lambda$ is the cut-off) in the presence of electron-electron (e-e)
interactions for a two-dimensional circular (three-dimensional spherical) 
Fermi surface \cite{Tsai1}. In particular, that work established 
that the stable FP of this problem is described by the 
Eliashberg-Migdal theory, which
contains BCS as its weak-coupling limit. Since it has 
few scattering channels \cite{Shankar}, the circular (spherical) 
Fermi surface, which is a consequence of the plane-wave character of the  electronic wavefunctions, is a particularly simple problem. This is not  the case for Fermi surfaces that derive from localized electronic orbitals  that can be either unstable to exotic SC (such as p- and d-wave) or  to density wave (charge and spin) phases.  
Arguably, 
the most famous example of this phenomenon are
the one-dimensional (1D)
conductors where the nesting associated with Fermi points leads to a richness
of phases that can be described by g-ology, bosonization
\cite{Solyom,vojt}, and the functional RG \cite{Tam}.  
However, the 1D problem is very particular because of the restricted 
phase space for interaction, and even in 1D, the treatment of the e-ph 
interaction is problematic \cite{vojt_phonons}. 

The recently developed e-ph  coupled functional renormalization group (FRG) \cite{Tsai1} includes the merits of the FRG
for the pure e-e interacting system
\cite{Shankar,Zanchi,Halboth} and goes beyond the two-step renormalization group for the e-ph  interacting system
\cite{Caron,Zimanyi,Seidel}, by allowing 
the systematic study of the full frequency dependence of couplings. In this paper we use this newly extended FRG
to study the two-leg ladder Holstein-Hubbard model that 
describes a quasi-1D system in the
presence of e-e and e-ph
interactions. The ladder system, consisting of two 1D chains coupled by  inter-chain hopping, can be considered as an intermediate between 1D and  2D systems, since it shares properties with both: it has a limited number of  scattering channels, as in 1D, but also can show SC with 2D characteristics,  such as d-wave order parameter \cite{White,Balents}. As a result, ladders have been considered by many as a theoretical testbed for the understanding of high-Tc superconductivity \cite{rice}.

Furthermore, the interest in the e-ph problem in high-Tc has been revitalized by experimental evidence that  the e-ph
coupling may play a critical role in these systems \cite{Lanzara,Shen}.  
Apart from the cuprates, 
ladder systems have also attracted much attention recently because, 
from the technical point of view, they are amenable to field 
theoretical calculations and to reliable numerical simulations, 
in contrast to
truly 2D systems. Studying the interplay between the e-e and e-ph
interactions in a ladder is also important for the understanding the 
physics of some low-dimensional materials, such as molecular crystals 
and charge transfer solids \cite{Organics}. 
 
 \begin{figure}[2band]
\centerline{
\includegraphics*[height=0.2\textheight,width=0.4\textwidth,viewport=68 215
510 550,clip]{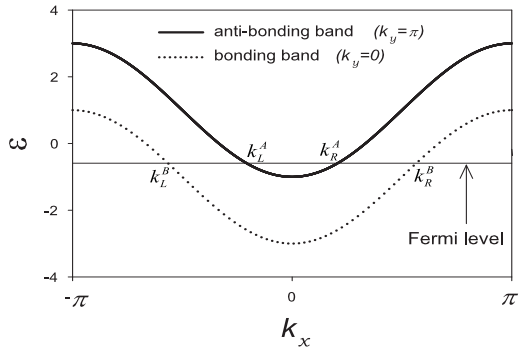}} \caption{Energy dispersion for free electrons in a
2-leg ladder.} 
\label{fig:n}
\end{figure}
 
The Holstein-Hubbard model (HHM) is described by the following Hamiltonian (we use units such that $\hbar =1=k_B$):  
\begin{eqnarray}
H &=& - t \sum_{\langle i,j \rangle,\sigma}(c_{i,\sigma}^{\dagger}c_{j,\sigma}
    + H.c.) + U\sum_{i}n_{i,\uparrow}n_{i,\downarrow}\nonumber\\
    & & + g_{\rm{ep}} \sum_{i,\sigma}(a_{i}^{\dagger}+a_{i})n_{i,\sigma} + \omega_0 \sum_{i} a_{i}^{\dagger}a_{i} \, ,
 \label{hhm}
\end{eqnarray}
where $c_{i,\sigma}^{\dagger}$ ($c_{i,\sigma}$) creates (annihilates) an 
electron at site $i$ with spin $\sigma$ ($n_{i\sigma}$ is the electron number 
operator),  $a_{i}^{\dagger}$ ($a_{i}$) creates (annihilates) an optical 
phonon at site $i$ with energy $\omega_0$, $t$ is the hopping energy of the 
electrons along legs and rungs of the ladder, $U$ is the on-site e-e
interaction, and $g_{\rm{ep}}$ is the e-ph
coupling energy. In the absence of e-e
($U=0$) and e-ph
($g_{\rm{ep}}=0$) interactions, the energy spectrum of the 
electrons has the form shown in Fig.~\ref{fig:n}. 
In what follows we use units such that $t=1$. 

One can reformulate the problem in terms of path integrals, and since the  phonon fields enter quadratically in the action, they 
can be integrated out exactly \cite{Tsai1}, generating an 
effective action $S$ for the electrons that can be written as:
\begin{eqnarray}
S &=& \int_{\underline{k},\sigma} \psi^{\dagger}_{\sigma}(\underline{k}) (i\omega - \epsilon_{k}) \psi_{\sigma}(\underline{k})   
\\ 
\nonumber 
&+& \int_{\{\underline{k}\},\sigma,\sigma^{'}} g(\underline{k_{1}},\underline{k_{2}},\underline{k_{3}},\underline{k_{4}}) \psi^{\dagger}_{\sigma}(\underline{k_{4}})\psi^{\dagger}_{\sigma^{'}}(\underline{k_{3}})\psi_{\sigma^{'}}(\underline{k_{2}})\psi_{\sigma}(\underline{k_{1}}) ,
\end{eqnarray}
where $\psi$ is the electron field, and the coupling functions for electron-electron interactions in the momentum space are given by: 
\begin{eqnarray}
g(\underline{k_1},\underline{k_2},\underline{k_3},\underline{k_4})
        = U -  2g_{\rm{ep}}^2\omega_0/[\omega_0^2 +(\omega_1 - \omega_4)^2],
\label{eq:g0}
\end{eqnarray}
where $\underline{k} = (k,\omega)$. Each coupling is a function of the
in-coming and out-going momenta and frequencies, subjected to the
conservation of momenta, $k_1+k_2=k_3+k_4$ and energy
$\omega_1+\omega_2=\omega_3+\omega_4$. In a purely e-e
interacting system, the coupling functions are always independent of the frequency. For a generic system with e-ph
interactions, the couplings become frequency dependent, i.e.,
retarded. We choose $g(\underline{k_1},\underline{k_2},\underline{k_3},\underline{k_4})$ to represent the
scattering between electrons with opposite spins. The coupling functions
for scattering between electrons with parallel spins are determined by
those with opposite spin due to the SU(2) spin symmetry. 

Following ref.~[\onlinecite{Tsai1}], our only assumption is that the full 
bandwidth
of the two-leg ladder system is larger than any of the interaction terms: 
i.e.,  that $6 t \gg U, g_{\rm{ep}},\omega_0$, leaving the relative values of $U$, $\omega_0$ and
$g_{\rm{ep}}$, free. The strong coupling limit of the electron-phonon problem occurs when $\lambda\equiv 2 N(0) g_{\rm{ep}}^2/\omega_0 \gg 1$. This regime  is accessible in our
functional RG approach. Utilizing the time-reversal symmetry, exchange symmetry, and reflection symmetry, the number of independent couplings in the Fermi surface for the 2-leg ladder at incommensurate fillings is 12 (each coupling is a function of three momenta and three frequencies, the frequency dependences are suppressed for clarity).
There are 6 intra-band couplings:
\begin{eqnarray}
g_{4}^{A} = g(k_{L}^{A},k_{L}^{A},k_{L}^{A},k_{L}^{A}),&g_{4}^{B} = g(k_{L}^{B},k_{L}^{B},k_{L}^{B},k_{L}^{B}),\nonumber \\
g_{1}^{A} = g(k_{L}^{A},k_{R}^{A},k_{L}^{A},k_{R}^{A}),&g_{1}^{B} = g(k_{L}^{B},k_{R}^{B},k_{L}^{B},k_{R}^{B}),\nonumber \\
g_{2}^{A} = g(k_{L}^{A},k_{R}^{A},k_{R}^{A},k_{L}^{A}),&g_{2}^{B} = g(k_{L}^{B},k_{R}^{B},k_{R}^{B},k_{L}^{B}); 
\end{eqnarray}
and 6 inter-band couplings:
\begin{eqnarray}
g_{1}^{C} = g(k_{L}^{A},k_{L}^{B},k_{L}^{A},k_{L}^{B}),&g_{2}^{C} = g(k_{L}^{A},k_{L}^{B},k_{L}^{B},k_{L}^{A}),\nonumber \\
g_{1}^{D} = g(k_{L}^{B},k_{R}^{B},k_{L}^{A},k_{R}^{A}),& g_{2}^{D} = g(k_{L}^{B},k_{R}^{B},k_{R}^{A},k_{L}^{A}),\nonumber \\
g_{1}^{E} = g(k_{L}^{B},k_{R}^{A},k_{L}^{B},k_{R}^{A}),&g_{2}^{E} = g(k_{L}^{B},k_{R}^{A},k_{R}^{A},k_{L}^{B}).
\end{eqnarray}

For pure e-e interactions the Fermi velocities are renormalized
by $g_{4}^{A}$ and $g_{4}^{B}$, and the renormalizations of scattering
between particles on the same side of the Brillouin zone at different bands,
$g_{1}^{C}$ and $g_{2}^{C}$, are absent in the leading order. This is not necessarily valid for couplings with frequency dependence. Moreover, these couplings generate self-energy corrections, which have been shown to be important for studying the system with retardation effects \cite{Tsai1}, leading to the renormalization of Fermi velocities and quasiparticle lifetime. Thus we keep these couplings even though they are not renormalized in the  non-retarded system . 
In leading order, our renormalization group equations for the coupling
function, $g_{\Lambda}(\underline{k}_1,\underline{k}_2,\underline{k}_3)$, and
for the self-energy, $\Sigma_{\Lambda}(\underline{k})$, are given by:
\begin{eqnarray}
\lefteqn{\partial_{\Lambda} g_{\Lambda}(\underline{k_1},\!\underline{k_2},\!\underline{k_3}) =}
\nonumber\\
&-&\!\!\!\!\int\!d\underline{p} \partial_{\Lambda}\!
[G_{\Lambda}(\underline{p})G_{\Lambda}(\underline{k})] g_{\Lambda}(\underline{k_1},\!\underline{k_2},\!\underline{k})
g_{\Lambda}(\underline{p},\!\underline{k},\!\underline{k_3})
\nonumber\\
&-&\!\!\!\!\int\!d\underline{p} \partial_{\Lambda}\!
 [G_{\Lambda}(\underline{p})G_{\Lambda}(\underline{q_1})]
g_{\Lambda}(\underline{p},\!\underline{k_2},\!\underline{q_1}) g_{\Lambda}(\underline{k_1},\!\underline{q_1},\!\underline{k_3})
\nonumber\\
&-&\!\!\!\!\int\!d\underline{p} \partial_{\Lambda}\!
 [G_{\Lambda}(\underline{p})G_{\Lambda}(\underline{q_2})][-\!2g_{\Lambda}(\underline{k_1},\!\underline{p},\!
 \underline{q_2})g_{\Lambda}(\underline{q_2},\!\underline{k_2},\!\underline{k_3})
\nonumber\\
&+&\!\!g_{\Lambda}\!(\underline{p},\!\underline{k_1},\!\underline{q_2})\! g_{\Lambda}\!(\underline{q_2},\!\underline{k_2},\!\underline{k_3})\!\!+\!g_{\Lambda}\!(\underline{k_1},\!\underline{p},\!\underline{q_2})
g_{\Lambda}\!(\underline{k_2},\!\underline{q_2},\!\underline{k_3})],
\end{eqnarray}
\begin{eqnarray}
\partial_{\Lambda} \Sigma_{\Lambda}(\underline{k}) \!=\! 
-\!\!\!\int\!d\underline{p} \partial_{\Lambda} [G_{\Lambda}(\underline{p})]
[2g_{\Lambda}(\underline{p},\underline{k},\underline{k})\!-\!g_{\Lambda}(\underline{k},\underline{p},\underline{k})],
\end{eqnarray}
where $\underline{k}=\underline{k_1}+\underline{k_2}-\underline{p}$, $\underline{q_1}=\underline{p}+\underline{k_3}-\underline{k_1}$, $\underline{q_2}=\underline{p}+\underline{k_3}-\underline{k_2}$,
$\int d\underline{p}=\int
dp\sum_{\omega}1/(2\pi\beta)$, and $G_{\Lambda}$ is the self-energy corrected propagator
at cutoff $\Lambda$. The initial condition for the
coupling functions is given by (\ref{eq:g0}). The self-energies are equal to zero at $\Lambda=\Lambda_0$.

For a non-retarded system the low energy instability can be obtained from the
RG flow of the couplings, and different phases can be identified by the FP corresponding to the different spin and charge modes. This procedure may not be appropriate for system with electron-phonon coupling. Numerical evidence indicates that the half-filled Holstein-Hubbard model violates the Luttinger relations \cite{Tezuka}. Therefore, instead of following the flows of the couplings, we explicitly construct the flows of the  susceptibilities of different order parameters 
to obtain a unbiased answer within the FRG.  The SC susceptibility, $\chi^{\delta}(k,\omega)$, is defined as:  
\begin{eqnarray}
\label{SC_susceptibility}
\chi^{\delta}_{\Lambda}(0,0)\!\!=\!\!\int\!\!D(1,2)
f^{\delta}(p_{1})f^{\delta}(p_{2})\langle
c_{p_{1},\downarrow}c_{-p_{1},\uparrow}c_{-p_{2},\uparrow}^{\dagger}c_{p_{2},\downarrow}^{\dagger}\rangle\!,
\end{eqnarray}
where $\delta=s,d$ for s- and d-wave SC, respectively \cite{com}.
The RG equations are: 
\begin{eqnarray} \label{SC_susceptibility RGE}
\partial_{\Lambda} \chi^{\delta}_{\Lambda}(0,0) = \int d\underline{p} \partial_{\Lambda}[
G_{\Lambda}(\underline{p})G_{\Lambda}(-\underline{p})](Z^{\delta}_{\Lambda}(\underline{p}))^2,\\
\partial_{\Lambda}\! Z^{\delta}_{\Lambda}\!(\underline{p})  \!\!=\!\!
-\!\!\!\int\!\!\!d\underline{p}^{\prime}\!\partial_{\Lambda}\![
G_{\Lambda}\!(\underline{p}^{\prime})G_{\Lambda}\!(\!-\underline{p}^{\prime})\!]\!Z^{\delta}_{\Lambda}\!(\underline{p^{\prime}})g_{\Lambda}\!(\underline{p^{\prime}}\!,\!\!-\underline{p^{\prime}},\!\!-\underline{p},\!\underline{p})\!.
\end{eqnarray}
The function $Z_{\Lambda}^{\delta}(\underline{p})$ is the effective vertex in the definition for the susceptibility $\chi_{\Lambda}^{\delta}$. Its initial condition, at $\Lambda=\Lambda_0$, is $1$ for s-wave pairing and $cos(k_x)-cos(k_y)$ for d-wave pairing. The RG equations for susceptibilities are solved with initial condition $\chi^{\delta}_{\Lambda=\Lambda_{0}}(0,0)=0$. The dominant
instability in the ground state is given by the most divergent susceptibility by solving the RG equations numerically. In the calculations reported here, we have discretized
the frequency axis into 9 points and have checked that our results are essentially unchanged when additional discretization points are added. 

Without e-ph coupling the d-wave pairing is the generic dominant instability for $0$ to $0.5$ doping (half- to quarter-filled), except in a narrow region  in the parameter space where all the charge and spin degrees of freedom are  gapless. The large region of d-wave pairing in the ladder comes from the  anisotropy between the
transverse and the parallel directions, which leads the pair scattering to develop a directional dependence when the high energy degrees of freedom are eliminated. For the Holstein coupling, the phonon-mediated attractive retarded interaction  is independent of the momentum transfer and is thus isotropic. The attractive interaction is relevant in generating pair-scattering, since it leads the particle-particle ladder diagrams to acquire logarithmic divergences. The d-wave pairing from repulsive interaction does not come from this divergence, and thus the attractive interaction is very effective in leading the pairing instability, and more importantly, this pairing has s-wave symmetry. Therefore, the e-ph coupling tends to enhance the s-wave component in the pairing scattering matrix, and once this coupling exceeds a critical value, the s-wave susceptibility overcomes the d-wave, causing a transition from d- to s-wave SC pairing symmetry.  This result can only  be seen in strong coupling and, hence, is not accessible to the two-step RG \cite{Seidel}. A study of the fully retarded problem for a 2D circular Fermi surface has similarly shown how the two-step RG breaks down in the strong coupling ($\lambda \gg 1$) limit \cite{Tsai2}.    

\begin{figure}[htb]
\centerline{
\includegraphics*[height=0.19\textheight,width=0.245\textwidth,viewport=60 210 500 550,clip]{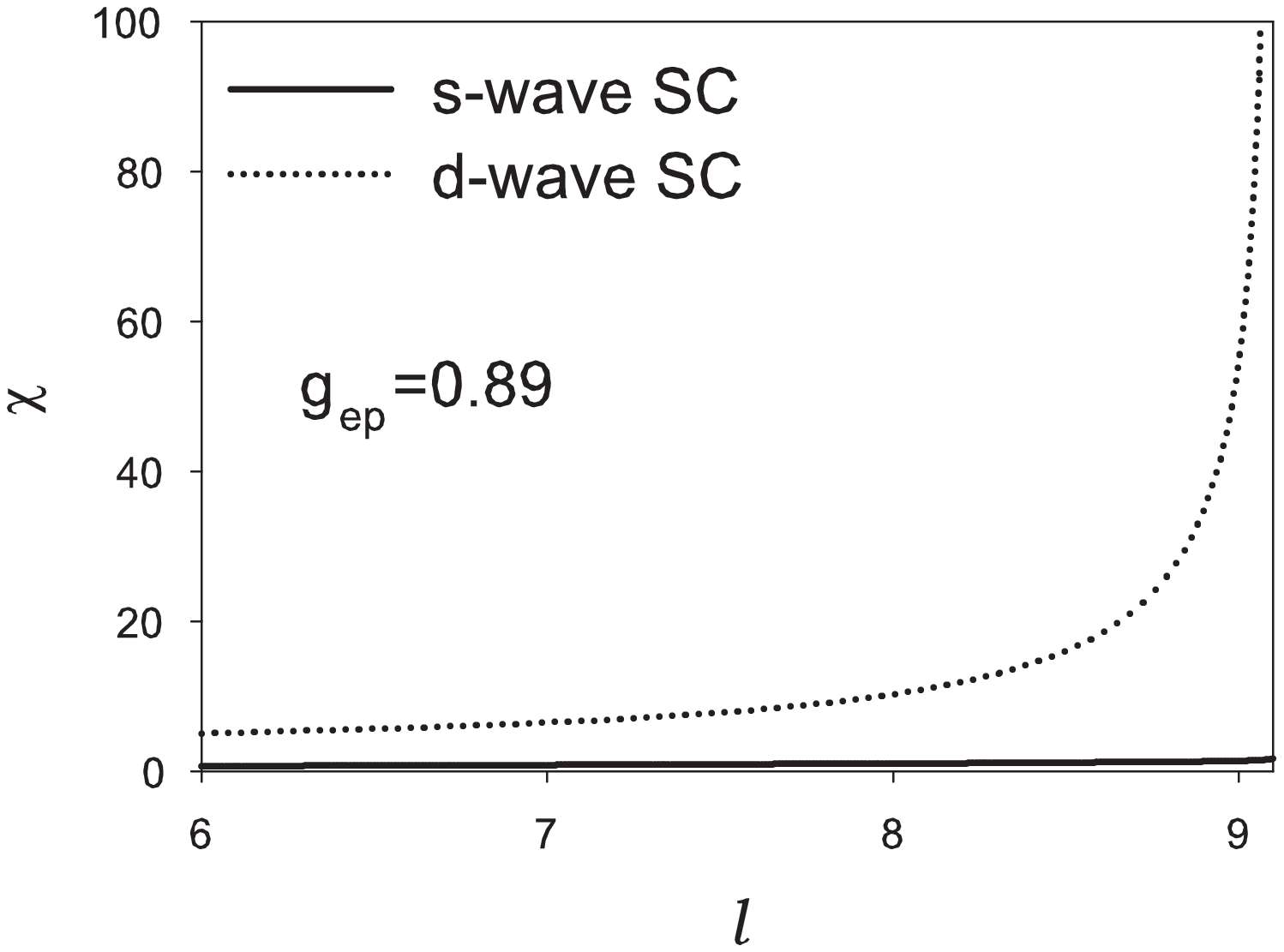}
\includegraphics*[height=0.19\textheight,width=0.245\textwidth,viewport=60 210 500 550, clip]{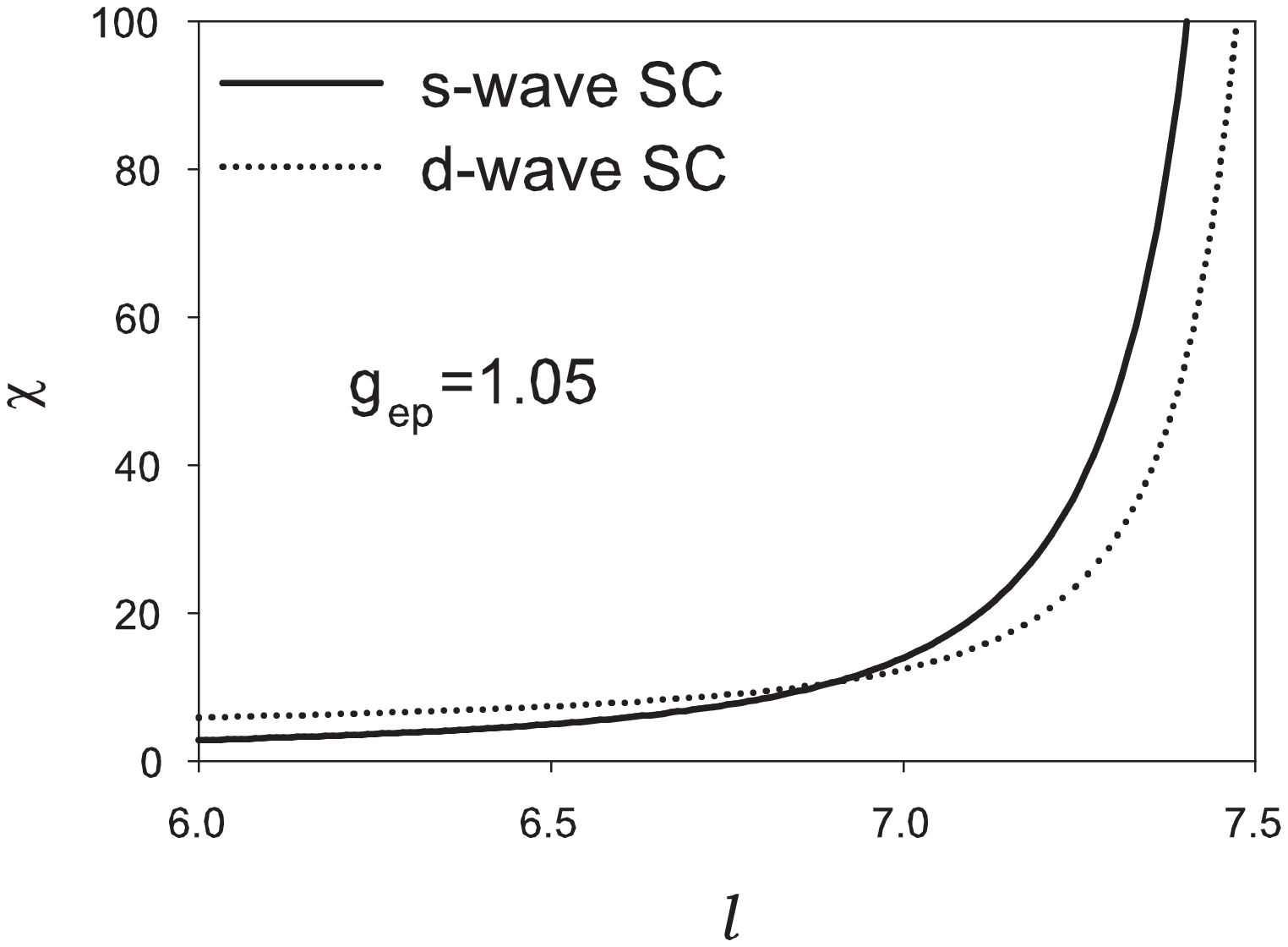}
}
\caption{The flows of s-wave SC (solid lines), and d-wave SC (dotted lines) susceptibilities as a
function of $l = \ln(\Lambda_0/\Lambda)$ for $U=2$, $\omega_0=1$, and doping
of $0.2$.} 
\label{fig:chi_flows}
\end{figure}

In Fig.~\ref{fig:chi_flows} we show a representative RG flow in which
we observe the change in the dominant susceptibility for increasing e-ph coupling $g_{\rm{ep}}$. Although the isotropic phonons tend to break the d-wave pairs, as long as the e-ph coupling is small compared to the e-e interactions, the d-wave pairing remains the dominant instability. However, the s-wave overcomes the d-wave pairing when the effective interaction in the anti-adiabatic limit, $U_{{\rm effective}} = U-2g_{\rm{ep}}^2/\omega_{0}$, is near
zero. As the doping increases, the critical values for the pairing transition remain close to the points where the anti-adiabatic effective interaction vanishes. Notice that for a ladder, the d-wave SC region {\it increases} with doping and is rather stable against s-wave near
quarter filling, because the Fermi surface becomes more anisotropic with doping. This result should be contrasted with the 2D case where d-wave superconductivity seems to be suppressed at larger doping because the Fermi surface becomes more isotropic. Hence, our results show that one has to be cautious when comparing results for ladders with the true 2D problem. For doping close to the quarter filling, we find noticeable changes in the critical values, and a substantially stronger electron-phonon interaction is needed to drive the s-wave pairing over the d-wave pairing. This can be understood from the fact that the difference between the Fermi velocities of
the anti-bonding and bonding bands becomes larger as the system is doped away from half-filling (see Fig.~\ref{fig:n}). From the perspective of the Fermi surface renormalization group, this difference leads to different contributions to pair scattering at $k_y=0$ and at $k_y=\pi$, which in turn lead to a phase difference in the pairing matrix elements which is manifested as a  d-wave pairing channel. This is the reason that the d-wave channel becomes stronger as the doping gets closer 
to quarter filling. Nevertheless, the s-wave remains still the dominant instability for sufficiently strong e-ph coupling. 

Exactly at quarter filling, the Fermi surface in the anti-bonding band
shrinks to a point, which leads to a van Hove singularity at the center of the anti-bonding band. Also, at this filling the distance between the Fermi points in the bonding band is commensurate, which results in the presence of an extra Umklapp scattering in the bonding band, $g_{3}^{B} = (k^B_L,k^B_L,k^B_R,k^B_R)$. We do not find that the Umklapp scattering is sufficient to drive the system to spin density wave (SDW) phase. In contrast  to the case without phonons $\lambda=0$ \cite{Balents} we do not find a Luttinger liquid phase (C2S2). The d-wave pairing is still the dominant instability for small e-ph coupling, and the s-wave again overcomes the d-wave at strong coupling.  Below quarter filling, the anti-bonding band is empty. Only two Fermi points remain on the bonding band. The system becomes essentially 1D as the effect of the anti-bonding band is then irrelevant. 

We have also investigated the problem exactly at half-filling. In this 
particular commensurate doping  there is a direct competition between a  SDW and a CDW phase. In the absence of e-ph interaction the SDW phase (with a charge gap but no spin gap) is very robust 
and is not destabilized at weak coupling. At intermediate coupling 
($\lambda \approx 1$), however. the SDW phase becomes unstable to a CDW phase with a charge gap. This result cannot be obtained in the weak coupling methods discussed previously. 

The final results of our study are summarized in the phase diagram of 
Fig.~\ref{phd}. We find that the e-ph coupling at intermediate and strong coupling leads to substantial qualitative differences from previous results, including  the appearance of new phases: CDW and s-wave SC. Moreover, we see that  the d-wave phase is stabilized close to quarter filling, in contrast with the 2D case.

In conclusion, we have studied the effect of e-ph coupling on the phase diagram of a two-leg ladder, using the Holstein-Hubbard model and applying the functional renormalization group method. This method deals with the retardation effects in a systematic way. In particular, when the e-ph interaction strength is comparable to the e-e
interactions and $\lambda \gg 1$, there are qualitative differences in the phase diagram from that predicted in previous studies. In particular, in addition to the  SDW and d-wave SC phases found previously, we establish that for  strong e-ph coupling, both CDW and s-wave SC can arise.

A.H.C.N. was supported through NSF grant DMR-0343790.


\begin{thebibliography}{20}
\bibitem{com} $\int D(1,2)\equiv \int_{|\xi_{1}|>\Lambda}d\xi_{1}J(\xi_{1})
\int_{|\xi_{2}|>\Lambda}d\xi_{2}J(\xi_{2})$, $J(\xi)$ is the Jacobian for the coordinate transformation from $k$ to $\xi_k$,  $f^s(p)=1$, and $f^{d}(p) = \cos(k_x) - \cos(k_y)$.
\end{thebibliography}
\end{document}